\documentclass{jpp}
\usepackage{graphicx}

\usepackage[utf8]{inputenc}
\usepackage[T1]{fontenc}
\usepackage{amsmath}
\usepackage{mathtools}
\usepackage{aas_macros}
\usepackage{layouts}
\usepackage{xcolor}
\usepackage{natbib}
\usepackage{appendix}
\usepackage{array}
\usepackage{ulem}

\newcommand{\pD}[2]{\frac{\partial #2}{\partial #1}}

\newcommand\bb[1]{\mbox{\boldmath{$#1$}}}

\newcommand{\unit}[1]{\hat{\bb{#1}}}

\newcommand{\rmd}{{\rm d}}

\newcommand{\const}{{\rm const}}

\newcommand{\vthp}{v_{{\rm thp}}}

\newcommand{\kprl}{k_\parallel}
\newcommand{\kresP}{k_{\parallel{\rm P}}}
\newcommand{\kresO}{k_{\parallel{\rm O}}}
\newcommand{\kprlp}{k_{\parallel{\rm P}}}

\newcommand{\kprp}{k_\perp}

\newcommand{\vprl}{v_\parallel}
\newcommand{\vprp}{v_\perp}
\newcommand{\vprlres}{v_{\parallel{\rm res}}}

\newcommand{\vresP}{v_{\parallel{\rm P}}}
\newcommand{\kprlres}{k_{\parallel{\rm res}}}
\newcommand{\vA}{v_{\rm A}}

\newcommand{\Rsun}{R_{\odot}}
\newcommand{\PICWsqrt}{\sqrt{1+4\left(\kprl d_{\rm p}\right)^{-2}}}

\newcommand{\omegatilde}{\tilde{\omega}}
\newcommand{\kprltilde}{\tilde{k}_{\parallel}}
\newcommand{\kprptilde}{\tilde{k}_{\perp}}

\newcommand{\omegaO}{\omega_{k\rm r,O}}
\newcommand{\omegaOprl}{\omega_{k_\parallel\rm r,O}}
\newcommand{\omegakr}{\omega_{k\rm r}}
\newcommand{\omegakprlr}{\omega_{k_\parallel\rm r}}
\newcommand{\omegaP}{\omega_{k_\parallel\rm r,P}}
\newcommand{\kprlturb}{k_{\parallel,{\rm CB}}}
\newcommand{\kprpturb}{k_{\perp,{\rm CB}}}

\shorttitle{Cyclotron breaking}
\shortauthor{E.~L.~Yerger, B.~D.~G.~Chandran, V.~David, T.~A.~Bowen, and S.~D.~Bale}

\title{Cyclotron breaking: a mechanism for parallel ion cyclotron waves to heat the fast solar wind}

\author{Evan L.~Yerger\aff{1}
  \corresp{\email{evan.yerger@unh.edu}},
  Benjamin D.~G.~Chandran\aff{1,2},
  Vincent David\aff{1,3},
  Trevor A.~Bowen\aff{4},
  Stuart D.~Bale\aff{4,5}
}

\affiliation{
\aff{1}Space Science Center, University of New Hampshire, Durham, NH 03824
\aff{2}Department of Physics, University of New Hampshire, Durham, NH 03824
\aff{3}Department of Physics and Astronomy, Dartmouth College, Hanover, NH 03755
\aff{4}Space Sciences Laboratory, University of California, Berkeley, CA 94720
\aff{5}Physics Department, University of California, Berkeley, CA 94720
}

\begin{document}

\maketitle

\begin{abstract}
The \textit{Parker Solar Probe} (\textit{PSP}) mission has observed near-continuous power in parallel ion cyclotron waves (PICWs) in the young, fast solar wind. These waves are unlikely to be directly produced by the turbulent cascade and are likely born of a local instability; yet, they are observed to both cool -- and heat -- the plasma. We propose that these observations can be self-consistently explained as the natural consequence of PICWs propagating in the inhomogeneous solar wind after they have been driven unstable. In this work, we argue that strong proton heating by a turbulent cascade of oblique ICWs will result in PICWs being driven unstable in a process known as quasi-linear focusing. Because the power in the turbulent cascade is concentrated at scales above the turbulent transition region, PICWs will be driven unstable within a range of wave numbers parallel to the background magnetic field, $\kprl$, that is bounded from above by~$\kprlp^*$, corresponding to the start of the transition region. As unstable PICWs propagate away from the sun to regions of lower proton density, their $\kprl$, multiplied by the proton inertial length~$d_{\rm p}$, increases. Eventually, the $\kprl d_{\rm p}$ of the PICWs becomes larger than $\kprlp^*d_{\rm p}$ and the waves damp, heating the solar wind. We call this effect `cyclotron breaking', in analogy with ocean waves breaking on the shore. We then discuss the testable predictions of the theory, including a distinct heating signature in which PICWs cool fast protons and heat slow protons at any given heliocentric distance~$r$. Finally, we conjecture that cyclotron breaking can lead to net heating by PICWs if the power emitted as PICWs decreases sufficiently rapidly with $r$ that local emission of PICWs is overwhelmed by the local damping of PICWs generated closer to the sun.
\end{abstract}

\section{Introduction}\label{sec:introduction}
The solar wind is a magnetized plasma that rapidly becomes super-sonic as it flows away from the sun and into the heliosphere. Some of the earliest evidence for the existence of the solar wind came from \citet{Biermann_1952}, who inferred a solar wind velocity ${U\sim 500-1500~{\rm km}~{\rm s}^{-1}}$ from observations of comet tails. These observations led to the discovery that a hot solar corona could accelerate the solar wind to super-sonic speeds via a thermal pressure gradient \citep{Parker_1958}. Since the first direct satellite measurements of the solar wind \citep{Gringauz_1960}, the solar wind has been characterized into two distinct phases based on velocity: fast $(U\gtrsim 700~{\rm km}~{\rm s}^{-1})$ and slow $(U\lesssim 400~{\rm km}~{\rm s}^{-1})$; though, there is growing evidence that Alfv{\'e}nicity, or correlation between fluctuations in the plasma velocity and magnetic field, and the composition of minor ions reveal at least two types of slow solar wind \citep[see][and references therin]{D'Amicis_2021}. Recent work has suggested that, past the Alfv{\'e}n critical point \citep[or, more realistically, Alfv{\'e}n surface; see][]{Badman_2025} -- where the solar wind becomes super-Alfv{\'e}nic -- the acceleration of the slow solar wind is best explained by gradients in the electron-enthalpy and heat fluxes and that of the fast solar wind is explained by gradients in the flux of energy carried by Alfv{\'e}n waves that are traveling away from the Sun \citep{Halekas_2023,Rivera_2025}.

The vast majority of the Alfv{\'e}n-wave flux is at scales that are much too large to directly heat the solar wind. Most likely, the transfer of wave energy from large to small scales is facilitated by a nonlinear cascade \citep{Coleman_1968,Velli_1989,Matthaeus_1999,Cranmer_vanBallegooijen_2005}. The most vigorous Alfv{\'e}n-wave cascades result from collisions between counter-propagating wave packets. Outward-propagating Alfv{\'e}n waves undergo a small amount of linear (non-WKB) reflection due to gradients in the background magnetic field and density \citep{Heinemann_Olbert_1980}. The presence of reflected waves is sufficient to initiate a nonlinear cascade that transports energy to small scales \citep{Dmitruk_2002}, where it can be dissipated \citep{Cranmer_vanBallegooijen_2003}. Indeed, a number of solar-wind models \citep{Cranmer_2007,Verdini_2010} have demonstrated that reflection-driven turbulence is sufficient to heat and accelerate the fast solar wind to observed velocities.

Despite the growing consensus that the cascade of anti-Sunward Alfv{\'e}n waves can power the fast solar wind, it is still not entirely clear what mechanisms are responsible for converting the energy in the turbulent fluctuations into particle heat and momentum. A complicating factor in discerning the relevant mechanisms is the fact that the solar wind is weakly collisional, with the mean free path for Coulomb collisions reaching the largest characteristic scales of the solar wind \citep{Verscharen_2019}. One clue towards resolving this mystery is the observation that proton heating dominates electron heating by ${4:1}$ in the fast solar wind \citep{Bandyopadhyay_2023}. When the energy in the background magnetic field dominates the thermal energy -- as it does in the inner heliosphere -- protons are generally understood to be collisionlessly heated through one of two observationally distinct mechanisms: resonant cyclotron heating by ion cyclotron waves (ICWs) \citep{Kennel_Engelmann_1966,Hollweg_Isenberg_2002,Isenberg_2011}, or via a non-resonant process called stochastic heating \citep{McChesney_1987,Chaston_2004,Chandran_2010,Bourouaine_Chandran_2013}\footnote{Although cyclotron and stochastic heating produce observationally distinct kinetic signatures, \citet{Johnston_2025} and \citet{Mallet_2026} have demonstrated that both mechanisms can be encapsulated under a unified theory.}. Recent work analyzing data from \textit{PSP} has provided observational evidence that ICW heating is an important mechanism for proton heating in the fast, super-Alfv{\'e}nic solar wind \citep{Bowen_2022,Shankarappa_2024}. 

Curiously, much of the ICW heating observed in the fast, super-Alfv{\'e}nic solar wind is found to be facilitated by parallel ICWs (PICWs), which are circularly polarized and propagate nearly parallel to the background magnetic field \citep{Liu_2023}. PICWs are pervasive in the inner heliosphere, appearing continuously for large segments of nearly every \textit{PSP} encounter \citep{Shankarappa_2025}. They are also observed in a narrow band of wave numbers parallel to the background magnetic field, $\kprl$, centered around $\kprl d_{\rm p}\sim 1$ \citep{Bowen_2024}. Here, $d_{\rm p}=\vA/\Omega_{\rm p}$ is the proton inertial length, $\vA=B/\sqrt{4\upi nm_{\rm p}}$ is the Alfv{\'e}n velocity, $\Omega_{\rm p}=q_{\rm p}B/m_{\rm  p}c$ is the proton gyrofrequency, $B$ is the magnitude of the background magnetic field, $c$ is the speed of light, and $n$, $q_{\rm p}$, and $m_{\rm p}$ are the proton number density, charge, and mass, respectively. Because PICWs are so separated in wave vector from the turbulent cascade of Alfv{\'e}n waves (that become oblique ICWs when $\kprl d_p\sim 1$), it is unlikely they can be generated from the cascade by nonlinear wave-wave interactions \citep{Zakarov_1992,Alexakis_2007}. A number of authors have proposed that PICWs might originate via quasi-linear focusing \citep{Chandran_2010_res,Isenberg_2011}. In quasi-linear focusing, heating by a turbulent cascade of oblique ICWs causes the proton velocity distribution to become so anisotropic that PICWs become unstable and grow, leading to a transfer of energy from oblique ICWs to PICWs through protons that are co-resonant with both waves. Quasi-linear focusing is also the most likely explanation for the emission of PICWs in hybrid-particle-in-cell simulations of imbalanced Alfv{\'e}nic turbulence \citep{Squire_2022,Zhang_2025}.

An outstanding issue with quasi-linear focusing explaining observations of PICWs is that the waves are often observed to heat the solar wind \citep{Bowen_2022,Bowen_2024}. Quasi-linear focusing is a process of PICW emission, not absorption, in which PICWs cool, rather than heat, the solar wind. The resolution of this issue is the primary focus of this work. We start with \S\ref{sec:picw_stability}, where we introduce the necessary background and machinery to understand quasi-linear focusing. This section includes the dispersion relations we use for both oblique ICWs and PICWs (\S\ref{sec:CPDR}), an introduction to the linear stability of ICWs in \S\ref{sec:linear_stability}, and an explanation of how the oblique ICW cascade can drive PICWs unstable in \S\ref{sec:ql_focusing}. With the necessary background established, we present the main result of this work, cyclotron breaking, in \S\ref{sec:cyclotron_breaking}. We begin \S\ref{sec:cyclotron_breaking} with a quick primer on the turbulent transition range (\S\ref{sec:transition_range_primer}), followed by a discussion on how the transition range affects quasi-linear focusing by limiting the band of wave numbers in which oblique ICW heating is strong and therefore PICWs are unstable (\S\ref{sec:ql_focusing_tr}). We then show that, as the resulting unstable PICWs propagate away from the Sun, they evolve so as to resonate with slower and slower protons. Eventually, PICWs will resonate with protons that are too slow to be resonantly heated by the turbulent cascade, at which point the PICWs will damp and heat the solar wind (\S\ref{sec:WKB_PICW}).

\section{PICWs driven unstable by the turbulent cascade}\label{sec:picw_stability}
In this section, we discuss how PICWs are driven unstable by an imbalanced turbulent cascade of oblique ICWs. It should be stated that the discussion here applies strictly for a homogeneous medium. However, in an inhomogeneous medium, corrections to the homogeneous dispersion relation will be $O(1/\kprl L)$, where $L$ is a characteristic length scale $|\rmd\ln B/\rmd r|^{-1}$ or $|\rmd \ln n/\rmd r|^{-1}$. Taking $\kprl d_{\rm p}\sim 1$, these corrections are at most $O(10^{-6})$ in the inner heliosphere and can therefore be safely ignored. In addition, we assume that the turbulent spectrum driving the PICWs consists of oblique ICWs (as defined in \S\ref{sec:CPDR}) and that it is imbalanced, i.e. that there is much more energy in waves propagating anti-sunward than sunward in the plasma rest frame. A fully self-consistent, hot-plasma treatment would make the following analysis analytically intractable, so we approximate the real part of the wave frequencies, $\omegakr$, using the cold-plasma dispersion relation and then use the quasi-linear analysis of \citet{Kennel_Wong_1967} to evaluate the sign of the imaginary part of the wave frequency,~$\gamma$. For a discussion of how the simple depiction of quasi-linear focusing depicted here is adapted to the solar wind, see \S\ref{sec:cyclotron_breaking}.

\subsection{Cold plasma dispersion relations for ion cyclotron waves}\label{sec:CPDR}
The real frequency $\omega_{k{\rm r}}$ of an ICW with wavenumbers $\kprl$ and $\kprp$, parallel and perpendicular to the background magnetic field, respectively, can be approximated by the cold plasma dispersion relation (equation 21 in chapter 2 of \citet{Stix_1992})
\begin{subequations}\label{eqn:wICW}
\begin{equation}
    \omegakr= \Omega_{\rm p}\sqrt{
        \dfrac{1}{2}\left(b-\sqrt{b^2-4c}\right)},
\end{equation}
where 
\begin{align}
    b&=\left(k d_{\rm p}\right)^2 + \left(\kprl d_{\rm p}\right)^2 + c,\\
    c &= k^2\kprl^2d_{\rm p}^4.
\end{align}
\end{subequations}
Here, $k=(\kprl^2+\kprp^2)^{1/2}$ is the magnitude of the wave number. If we now set $\kprp=0$, we obtain the PICW dispersion relation \citep{Hollweg_Isenberg_2002}
\begin{equation}\label{eqn:wPICW}
    \omegaP =\dfrac{\Omega_{\rm p}\left(\kprl d_{\rm p}\right)^2}{2}\left[\PICWsqrt-1\right].
\end{equation}

\subsection{ICW stability}\label{sec:linear_stability}
The linear stability of a plasma to a given wave mode is entirely due to the preferred direction of energy exchange between the wave mode and particle population, as mediated by resonant wave-particle interactions. Waves and particles, here cyclotron waves and protons, interact resonantly when $\omega_{k\rm r}$, as measured in a frame moving along the background magnetic field at the parallel velocity of the proton~$\vprl$, is an integer ($n$) multiple of the proton cyclotron frequency $\Omega_{\rm p}$ \citep{Stix_1992}, i.e. 
\begin{equation}\label{eqn:resonance}
    \omega_{k\rm r}-\kprl\vprl=n\Omega_{\rm p}.
\end{equation}
Here, $\vprl$ is measured in the frame traveling at the solar wind velocity $U$. One can either solve \eqref{eqn:resonance} for $\kprlres=\kprlres(\vprl,\kprp,n)$, the $\kprl$ of a wave with given $\kprp$ whose $n^{\rm th}$ cyclotron harmonic is resonant with a particle traveling at $\vprl$, or for
\begin{equation}\label{eqn:vprlres}
    \vprlres(\kprl,\kprp,n) = \dfrac{\omega_{k\rm r}-n\Omega_{\rm p}}{\kprl},
\end{equation}
the velocity of a particle that is resonant with the $n^{\rm th}$ cyclotron harmonic of a wave with a given $\kprl$ and $\kprp$.

For $n=0$, $\vprlres = \omega_{k\rm r}/\kprl$, i.e. the parallel resonant velocity for a given wave is its phase speed. The $n=0$ resonance is often referred to as the Landau resonance. Landau-resonant wave-particle interactions can be mediated by the component of the electric field parallel to the magnetic field, as is the case with Landau damping, or by the electric field components perpendicular to the magnetic field, as is the case with transit-time damping \citep{Stix_1992}. Qualitatively, if the derivative of the proton velocity distribution function, $f_{\rm p}$, with respect to $\vprl$ has the same sign as the wave phase velocity, $\omega_{k\rm r}/\kprl$, in the immediate vicinity of the resonant velocity $\vprl=\omega_{k\rm r}/\kprl$, more protons will give energy to the wave than take away from it. The net effect will be wave growth -- instability. Conversely, wave energy will be absorbed (damped) if the opposite is true. 

When $n\ne 0$, the wave-particle interactions are said to be cyclotron resonant. Consider a single plane wave. In the `wave frame,' which moves at the wave phase velocity along the background magnetic field~$\unit{b} \omega/\kprl$ (here $\unit{b}=\bb{B}/B$ is the unit vector along the magnetic field), the partial time derivative of the wave fields vanishes, and the total particle energy~$H = \frac{1}{2} mv^2 + q_{\rm p} \Phi$ is constant, where $\Phi$ is the electrostatic potential. For small-amplitude waves, $\Phi$ is a constant plus a small sinusoidal oscillation. Therefore, the particle kinetic energy~$\frac{1}{2} m_{\rm p} v^2$ undergoes only small oscillations, cannot change secularly, and is effectively conserved. Cyclotron-resonant wave-particle interactions can, however, cause particles to diffuse in the $v_\parallel - v_\perp $ plane along contours of constant energy in the wave frame, which are defined by the equation \citep{Kennel_Engelmann_1966}
\begin{equation}\label{eqn:ql_contours}
    \left(\vprl-\left.\dfrac{\omega_{k\rm r}}{\kprl}\right|_{\kprlres}
    \right)^2 + \vprp^2 = \const.
\end{equation}
The solutions to \eqref{eqn:ql_contours} are called quasi-linear contours. The intuition behind the stability of cyclotron-resonant waves (like ion cyclotron and whistler waves) is much the same as waves mediated by the Landau resonance. However, the relevant velocity-space derivative of $f_{\rm p}$ is given by the differential operator \citep{Kennel_Wong_1967}
\begin{equation}\label{eqn:KWG}
    G=\left(\left.\dfrac{\omega_{k\rm r}}{\kprl}\right.
    -\vprl\right)\pD{\vprp}{}+\vprp\pD{\vprl}{},
\end{equation}
which calculates the derivative of its operand along the quasi-linear contours \eqref{eqn:ql_contours}. To determine the stability of a wave mode, one must consider the net effect of all the resonant particles. Indeed, this consideration is taken into account in the full calculation of the linear growth rate in \citet{Kennel_Wong_1967}. For PICWs, one can show that the growth rate $\gamma_{\rm P}(\kprl)$ is positive when 
\begin{equation}\label{eqn:PICW_stability}
    \gamma_{\rm P}(\kprl)\propto
    {\rm sign}\left(\left.\dfrac{\omegaP}{\kprl}\right.
    \right)\int_0^\infty\rmd\vprp\,\vprp^2\left.\left[G_{\rm P}f_{\rm p}(\vprl,\vprp)\right]\right|_{\vresP} > 0.
\end{equation}
Here, $G_{\rm P}$ is the differential operator~$G$ of~\eqref{eqn:KWG} evaluated for $\omegakr=\omegaP$, and ${\vresP=v_{\parallel{\rm res}}(\kprl,0,1)}$ is the resonant parallel velocity for a PICW with parallel wave number $\kprl$.

PICWs are left circularly polarized and therefore only able to interact with protons via the $n=1$ cyclotron resonance \citep{Stix_1992}. It then follows that each $v_\parallel$ can resonate with only a single~$k_\parallel$, and there is a unique quasilinear contour passing through each point in velocity space \citep{Isenberg_2001}. This can be seen graphically with the aid of figure~\ref{fig:ql_focusing}(c). Technically, protons with a given~$v_\parallel$ whose magnitude is~$\lesssim v_{\rm A}$ could interact with oblique ICWs at multiple values of~$k_\parallel$ via different resonances~(i.e., $n\neq 1$). However, the values of $k_\parallel$ for~$n\neq 1$ are far outside the inertial range, and so we neglect them. Thus, effectively, when protons interact with oblique ICWs, the protons  at each~$v_\parallel$ interact with only a single~$k_\parallel$, and there is again a unique quasilinear contour passing through each point in the $v_\parallel - v_\perp$ plane. 

\subsection{Quasi-linear focusing}\label{sec:ql_focusing}
Quasi-linear focusing \citep{Chandran_2010_res, Isenberg_2011} is a process in which oblique ICWs from the turbulent cascade are `focused' into PICWs via wave-particle interactions. To explain this process, we assume for simplicity that all the oblique ICWs propagate with a $\kprp$ that is an a-priori known function of $\kprl$, $\kprpturb(\kprl)$, given by critical balance \citep{Goldreich_Sridhar_1995,Maron_Goldreich_2001,Sioulas_2025} and are present at all~${k_\parallel d_{\rm p}\lesssim 1}$. $\kprpturb(\kprl)$ is further discussed in \S\ref{sec:transition_range_primer}; however, it should be stated here that the results of this section -- and this work in general -- are insensitive to the exact functional dependence of $\kprpturb$ on $\kprl$. Let us define the real frequency of the oblique ICW with this choice of $\kprp$ to be ${\omegaOprl\equiv\omegakr(\kprl,\kprpturb(\kprl))}$. As these ICWs heat an initially Maxwellian plasma, the plasma will eventually reach a state, denoted~$f_{\rm p, O}$, of marginal stability to the ICWs, in which \citep{Kennel_Engelmann_1966,Isenberg_2001,Isenberg_2011}
\begin{equation}\label{eqn:fpO_defn}
    G_{\rm O}|_{\kresO}
    f_{\rm p,O}(\vprl,\vprp)=0.
\end{equation}
Here, $G_{\rm O}$ is the differential operator~$G$ of~(\ref{eqn:KWG}), evaluated for $\omegakr=\omegaOprl$. So that \eqref{eqn:fpO_defn} is purely a function of velocity, $G_{\rm O}$ is evaluated at the resonant parallel wave number for oblique ICWs, $\kresO\equiv k_{\parallel{\rm res,O}}(\vprl)$. After $f_{\rm p}$ reaches the state~$f_{\rm p,O}$, plasma heating by oblique ICWs ceases, and the oblique ICWs no longer damp. Moreover, $f_{\rm p,O}$, which is marginally stable to oblique ICWs, is unstable to PICWs, as we now show. Henceforth, we shall assume that both oblique ICWs and PICWs propagate anti-Sunward, so ${{\rm sign}\left(\omegaOprl/\kprl\right)={\rm sign}\left(\omegaP/\kprl\right)}=1$. The condition \eqref{eqn:PICW_stability} can be simplified by adding and subtracting the phase speed of the co-resonant oblique ICW, $(\omegaOprl/\kprl)|_{\kresO}$, {\em viz.} 
\begin{equation}\label{eqn:G_P}
\begin{split}
    G_{\rm P}f_{\rm p,O}(\vprl,\vprp)&=\left[\left(\dfrac{\omegaP}{\kprl}-\left.\dfrac{\omegaOprl}{\kprl}\right|_{\kresO}\right)\pD{\vprp}{}+G_{\rm O}|_{\kresO}\right]
    f_{\rm p,O}(\vprl,\vprp)\\
    &=\left(\dfrac{\omegaP}{\kprl}-\left.\dfrac{\omegaOprl}{\kprl}\right|_{\kresO}\right)\pD{\vprp}{}f_{\rm p,O}(\vprl,\vprp),
\end{split}
\end{equation}
where the last equality follows from the definition of $f_{\rm p,O}$ \eqref{eqn:fpO_defn}. With \eqref{eqn:G_P} in hand, we can integrate \eqref{eqn:PICW_stability} by parts. The remaining integral is is positive-definite, so the inequality reduces to ${\gamma_{\rm P}(\kprl)\propto \left.(\omegaOprl/\kprl)\right|_{\kresO(\vresP)}-\omegaP/\kprl>0}$.
Thus, a PICW will be unstable if the oblique ICW that is co-resonant, or resonates with protons of the same $\vprl$, has a faster phase velocity. 

\begin{figure}
    \centering
    \includegraphics{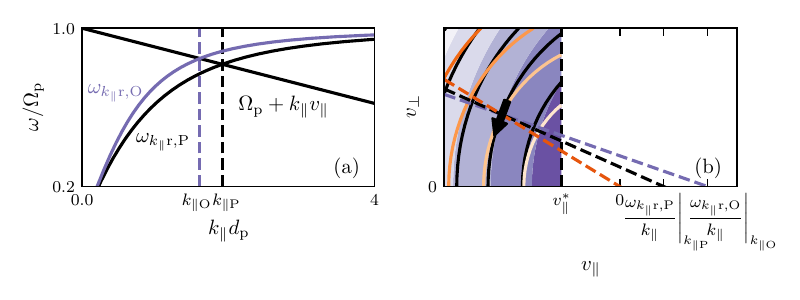}
    \caption{The frequency of an oblique ICW, $\omegaOprl$, here with (a very exaggerated value of) $\kprp d_{\rm p}=10$, is always larger than $\omegaP$ for a given $\kprl$. As a result, $\kresO>\kresP$, as shown by the intersection of $\omegaO$ and $\omegaP$ with $\Omega_{\rm p}+\kprl\vprl$ in (a). One can then show that ${(\omegaOprl/\kprl)|_{\kresO}-(\omegaP/\kprl)|_{\kresP}>0}$ (see text), a necessary condition for quasi-linear focusing. For this condition to be sufficient, the proton velocity distribution must be nearly constant along oblique ICW contours. In subfigure (b) (after figure 4 in \citet{Chandran_2010_res}), we show the proton velocity distribution function (darker purple denote contours of higher particle density) only for $\vprl<\vprl^*$, where $\vprl^*$ is the resonant parallel velocity associated with the start of the turbulent transition region (see \S\ref{sec:ql_focusing_tr}). Physically, PICWs provide alternate quasi-linear contours (black semi-circles) along which protons can diffuse. Because ${(\omegaOprl/\kprl)|_{\kresO}-(\omegaP/\kprl)|_{\kresP}>0}$ (the difference in subfigure b is exaggerated for clarity), protons will on average diffuse toward regions of velocity space with lower proton density, i.e. in the direction given by the black arrow. These regions of low proton density also have lower energy (contours of larger constant energy are given by darker orange semi-circles); the energy lost by protons is given to the PICWs, causing their amplitude to grow \citep{Chandran_2010_res}. Purple, black, and orange dashed lines, corresponding to the center of the semi-circular quasi-linear contours for oblique ICWs, PICWs, and contour of constant energy, respectively, are included to aid the eye.}
    \label{fig:qlf_explanation}
\end{figure}

One can equivalently view $\gamma_{\rm P}(\kprl)$ as a function of $\vprl$ by evaluating at $\kresP=k_{\parallel{\rm res,P}}(\vprl)$: 
\begin{equation}\label{eqn:qlf_vph_condition2}
    \gamma_{\rm P}(\kresP(\vprl))\propto \left.\dfrac{\omegaOprl}{\kprl}\right|_{\kresO}-\left.\dfrac{\omegaP}{\kprl}\right|_{\kresP}>0.
\end{equation}
In \eqref{eqn:qlf_vph_condition2}, as before, the growth rate of the resonant PICW is positive if the phase speed of the co-resonant oblique ICW is larger than that of the PICW. This condition was proven to be satisfied up to first-order in an asymptotic expansion in $|\vprl|/\vA$ in \citet{Chandran_2010_res}. In what follows, we show that the condition is satisfied in general for the cold-plasma dispersion relation. 
In Appendix~\ref{apx:proof}, we show algebraically that $\omegakr(\kprl,\kprp)>\omegakprlr$ for all $(\kprl,\kprp)>(0,0)$. This fact is also illustrated by graphs of $\omegaP$ and $\omegaOprl$ with a chosen value $\kprp$, as shown in figure~\ref{fig:qlf_explanation}(a). It is therefore true that for any $\kprl$, $\omegaOprl/\kprl>\omegaP/\kprl$, and that for any $\vprl$, $\kresP>\kresO$ (as shown in figure~\ref{fig:qlf_explanation}(a)). Because $\omegaP/\kprl$ is a monotonically decreasing function of $\kprl$, ${(\omegaOprl/\kprl)|_{\kresO}>(\omegaP/\kprl)|_{\kresO}>(\omegaP/\kprl)|_{\kresP}}$ and $\gamma_{\rm P}(\kresP(\vprl))>0$. 

\citet{Chandran_2010_res} also showed that, physically, PICWs can be driven unstable by strong oblique ICW heating because co-resonant PICWs provide alternate contours for protons to diffuse along. As the condition \eqref{eqn:qlf_vph_condition2} is generally satisfied, PICW contours will be shallower than oblique ICW contours at a given $\vprl$. There will therefore be a proton velocity-space-density gradient along PICW contours in the direction of smaller parallel speed. Protons will then naturally diffuse in the direction of decreasing particle density along the PICW contours in the direction increasing parallel speed. However, PICW contours are steeper than contours of constant energy, so the net diffusion of protons is towards velocity-space regions of lower kinetic energy. Total energy is conserved, so the kinetic energy lost by protons is resonantly transferred to the electromagnetic field as PICWs, driving the waves unstable. We show a highly simplified version of this process in figure~\ref{fig:qlf_explanation}(b).

\section{Cyclotron breaking}\label{sec:cyclotron_breaking}
In \S\ref{sec:picw_stability} we provided a general explanation of quasi-linear focusing that applied to a homogeneous plasma. Here, we extend that discussion to the solar wind via the addition of two important concepts. First, in \S\ref{sec:transition_range_primer}, we introduce the turbulent transition range. Then, in \S\ref{sec:ql_focusing_tr}, we argue that the drop in spectral power near the transition range constrains where quasi-linear focusing is active and where it is not. Finally, in \S\ref{sec:WKB_PICW}, we leverage properties of the trajectories of anti-Sunward PICWs in the inhomogeneous solar wind to argue that if PICWs are anywhere driven unstable, they must propagate to larger radii and damp, heating the solar wind. This implies a continuous process of PICW heating and cooling, which we call `cyclotron breaking'. 

\subsection{The turbulent transition range and its associated scales}\label{sec:transition_range_primer}

\begin{figure}
    \centering
    \includegraphics{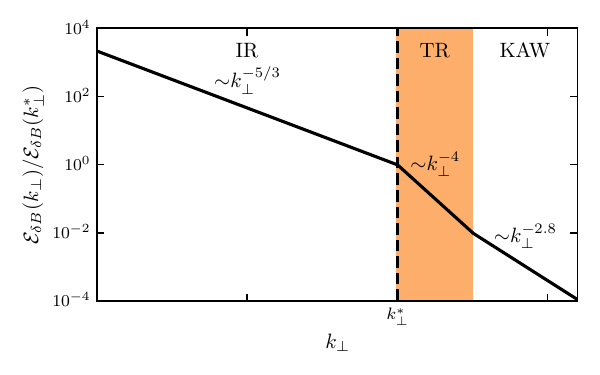}
    \caption{A reduced magnetic fluctuation spectrum $\mathcal{E}_{\delta B}(\kprp)$, typical of the solar wind. The transition range (TR), characterized by a steep ${\sim}\kprp^{-4}$ spectral slope, is sandwiched between the inertial range (IR) and kinetic-Alfv{\'e}n-wave range (KAW). We denote the perpendicular wavenumber at the boundary between the IR and TR as $\kprp^*$.}
    \label{fig:TR}
\end{figure}

The turbulent transition range is an interval of the turbulent spectrum in which the spectral scaling is observed to be very steep. While reports of the spectral index of the transition range vary, it is usually described as having a mean spectral index ${\sim}\kprp^{-4}$ \citep{Leamon_1998,Bowen_2022,McIntyre_2025}. This steep scaling is observed near $\kprp\rho_{\rm p}\sim 1$, where $\rho_{\rm p}=\vthp/\Omega_{\rm p}$ is the thermal proton gyroradius and $\vthp$ is the proton thermal speed. The transition range is sandwiched by the ${\sim}\kprp^{-3/2}$ or ${\sim}\kprp^{-5/3}$ inertial range at larger scales and the ${\sim}\kprp^{-2.8}$ kinetic-Alfv{\'e}n-wave range at smaller scales \citep{Alexandrova_2009,Podesta_2013}. In keeping with recent work \citep{Meyrand_2021}, we denote the perpendicular wave number at the boundary between the inertial range and transition range~$\kprp^*$, as illustrated in figure~\ref{fig:TR}. 

We assume that the 2D power spectrum of the fluctuating magnetic field~$\mathcal{E}_{\delta B}(k_\perp, k_\parallel)$, as reported in \citet{Squire_2022}, is a constant function of~$k_\parallel$ between $k_\parallel=0$ and a nonzero parallel wavenumber given by the critical-balance cone, $\kprlturb(\kprp)$. The inverse of this function, $\kprpturb(\kprl)$, is used to compute $\omegaOprl$, as discussed in \S\ref{sec:ql_focusing}. We take $\kprlturb(\kprp)$ to be an increasing function of $k_\perp$. The bulk of the fluctuation energy at each~$k_\perp$ then has a characteristic parallel wavenumber~${\sim}\kprlturb(\kprp)$, even though smaller parallel wavenumbers are also present. Therefore, the characteristic parallel wavenumber at the start of the transition range, $\kprl^* \equiv \kprlturb(\kprp^*)$, is the largest parallel wavenumber reached by the energy cascade within the inertial range. A depiction of $\mathcal{E}_{\delta B}(\kprp, \kprl)$ that includes $\kprp^*$, $\kprl^*$, and the function~$\kprlturb(\kprp)$ is shown in figure~\ref{fig:ql_focusing}(a). 

\subsection{Quasi-linear focusing near the turbulent transition range}\label{sec:ql_focusing_tr}
\begin{figure}
    \centering
    \includegraphics[]{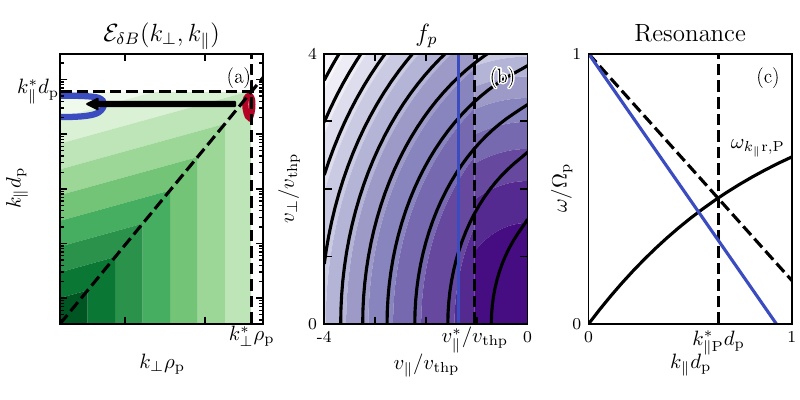}
    \caption{A diagrammatic description of quasi-linear focusing in the presence of the turbulent transition range. Our model turbulent spectrum for $\delta B$ is shown in (a), with darker green denoting larger values. The perpendicular and parallel start of the transition range, $\kprp^*$ and $\kprl^*$, and the critical balance anisotropy relationship $\kprlturb(\kprp)$, are given by the vertical, horizontal, and slanted dashed black lines, respectively. Heating by the cascade (red circle in (a)) pushes the proton velocity distribution $f_{\rm p}$, shown in (b), constant along oblique ICW quasi-linear contours for $\vprl < \vprl^*$, where $\vprl^*$ is the resonant parallel velocity for an oblique ICW with $\kprl=\kprl^*$ and $\kprp=\kprp^*$. Here, $f_{\rm p}$ is unstable to PICWs because oblique ICW quasi-linear contours are steeper than PICW contours (given by black lines in (b)). PICWs will therefore be emitted, cooling $f_{\rm p}$, in the wave number region given by the blue circle in (a). For $\vprl>\vprl^*$, $f_{\rm p}$ is much more isotropic because heating by oblique ICWs is much weaker there. Consequently, the contours of $f_{\rm p}$ are shallower than those of the PICWs, and the PICWs are damped. A given unstable PICW, resonating with protons with a $\vprl$ given by the vertical blue line in (b), will necessarily have $\kprl < \kprlp^*\simeq \kprl^*$. The PICW resonance diagram, (c), shows $\omegaP$ (black curved line) and the right-hand side of \eqref{eqn:res_PICW} (straight lines) -- the intersection of which is the solution to the resonance condition. Wave-particle interactions with faster protons constitute steeper lines and therefore smaller resonant $\kprl$. We approximate the extent of the excited spectrum of PICWs by the blue circle in (a).}
    \label{fig:ql_focusing}
\end{figure}

Let us ignore for a moment the expansion of the solar wind and focus on how the presence of the turbulent transition range bounds the regions of wave vector- and velocity-space where wave-particle interactions lead to PICW emission via quasi-linear focusing. Relative to the inertial range, there is vanishingly little power in the transition range. Indeed, observations of imbalanced turbulence by \textit{PSP} have shown that the spectral energy across the transition range drops by at least a factor of $100$ \citep{Bowen_2022,Bowen_2024b}. Let us assume for the sake of simplicity that the power in the oblique ICW spectrum falls to zero for $\kprp>\kprp^*$ and $\kprl>\kprl^*$. The highest-frequency oblique ICWs in the inertial range, depicted by the red circle in figure~\ref{fig:ql_focusing}(a), dominate the oblique-ICW heating rate.

The negligible power in oblique ICWs at $\kprl \gtrsim \kprl^*$ restricts strong cyclotron-resonant interactions between protons and oblique ICWs to $\vprl\lesssim\vprl^*$, where
\begin{equation}\label{eqn:vprlstar}
    \vprl^*=v_{\parallel{\rm res,O}}(\kprl^*,
    1)=\dfrac{\omegaOprl(\kprl^*)-\Omega_{\rm p}}{\kprl^*}
\end{equation}
is the solution to \eqref{eqn:vprlres} for oblique ICWs at the start of the transition range. We note that $\vprl^* < 0$, so faster protons moving in the direction opposite the wave-propagation direction can resonate, but slower protons cannot.
We plot $\vprl^*$ with a vertical dashed black line over a model proton velocity distribution in figure~\ref{fig:ql_focusing}(b). 
Let us assume the resonant heating by the oblique cascade is sufficiently strong to flatten the proton distribution constant along the quasilinear contours of \eqref{eqn:ql_contours} for oblique ICWs at $v_\parallel < v_\parallel^\ast$. For $\vprl>\vprl^*$, we assume the proton distribution has not been extensively heated by oblique ICWs and is significantly more isotropic than for $\vprl < \vprl^*$. We depict the resulting distribution in figure~\ref{fig:ql_focusing}(b). As discussed in \S\ref{sec:ql_focusing}, the proton distribution for $\vprl < \vprl^*$ is resonantly unstable to PICWs.

The resonance condition \eqref{eqn:resonance}, now applied to the relevant $n=1$ cyclotron resonance for PICWs, is
\begin{equation}\label{eqn:res_PICW}
    \omegaP(\kprl)=\kprl\vprl+\Omega_{\rm p}.
\end{equation}
Solutions to \eqref{eqn:res_PICW} are easily found pictorially in a resonance diagram, which we have included in figure~\ref{fig:ql_focusing}(c). We plot the PICW wave frequency $\omegaP$, the left-hand side of \eqref{eqn:res_PICW}, versus $\kprl$ with a solid black line. The right-hand side of \eqref{eqn:res_PICW} is a straight line originating from the top-left corner of the plot. As discussed in \S\ref{sec:ql_focusing}, the point at which the two lines intersect is a solution of the resonance condition. We have included two lines for two different parallel velocities. The slanted dashed line is $\kprl\vprl^*+\Omega_{\rm p}$, which intersects $\omegaP$ at
\begin{equation}\label{eqn:kprlPstar}
    \kprlp^*=\kresP(\vprl^*),
\end{equation} 
the parallel wave number for PICWs resonant with protons with $\vprl=\vprl^*$. We illustrate the value of $\kprlp^*$ with a black dashed vertical line in figure~\ref{fig:ql_focusing}(c). The blue solid line in figure~\ref{fig:ql_focusing}(c) is $\kprl\vprl+\Omega_{\rm p}$ for a parallel velocity $\vprl<\vprl^*$, which is itself shown by the vertical solid blue line in figure~\ref{fig:ql_focusing}(b). The blue solid line in figure~\ref{fig:ql_focusing}(c) intersects the wave frequency at a parallel wave number $\kprl<\kprlp^*$, where PICWs are unstable. This blue solid line is colored this way to represent an unstable PICW actively cooling the plasma. Although our discussion so far has focused on a single wave mode, we expect PICW emission will occur in a band of wave numbers at $\kprl\lesssim \kprlp^*$, as depicted by the blue circle in figure~\ref{fig:ql_focusing}(a). 

What about PICWs with $\kprl>\kprlp^*$? In our resonance diagram, these waves would be represented by lines that are shallower than the black dashed line. Shallower lines correspond to resonances with protons that have ${\vprl > \vprl^*}$. Thus, PICWs with ${\kprl>\kprlp^*}$ are strongly damped by the proton distribution. The result of quasi-linear focusing is therefore very little wave energy for $\kprl>\kprlp^*$ and a continuous buildup of PICW wave energy for $\kprl<\kprlp^*$. In a homogeneous plasma, there is no linear mechanism by which the wave energy at $\kprl<\kprlp^*$ can be transported to $\kprl>\kprlp^*$, where the wave energy can damp and heat the plasma. As it turns out, such a transport mechanism is the natural consequence of the WKB propagation of PICWs in the inhomogeneous solar wind. 

\subsection{WKB evolution of PICWs}\label{sec:WKB_PICW}
\begin{figure}
    \centering
    \includegraphics[]{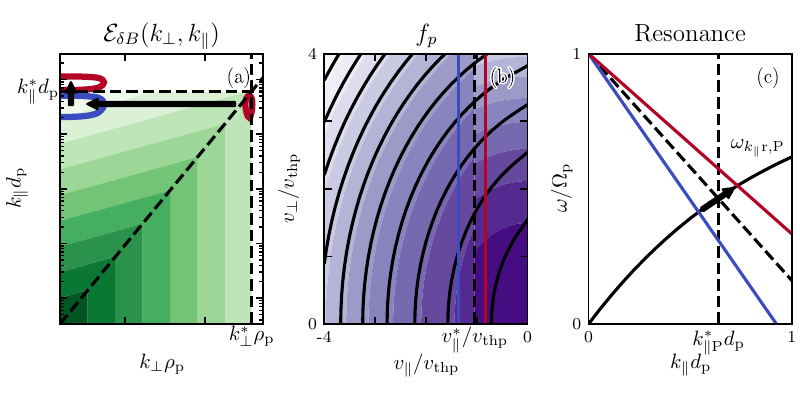}
    \caption{PICWs driven unstable by quasi-linear focusing in the vicinity of the turbulent transition range cool the solar wind (see figure~\ref{fig:ql_focusing}). PICWs that are at one radius unstable will propagate away from the Sun. As they do, the wave frequency and parallel wave vector will evolve according to \eqref{eqn:PICW_WKB}, i.e. along the (properly normalized) PICW dispersion relation towards larger values, as shown by the arrow in (c). The solution to the resonance condition changes appropriately, with PICWs resonating with slower and slower protons as they propagate -- shown by the red line in (b). At some point in their anti-Sunward trajectory, the PICWs will begin to resonate with the $\vprl>\vprl^*$ part of the proton velocity distribution (which has not been heated by the ICW cascade) and will be damped. The damped waves heat the solar wind, as emphasized by the red lines in (b) and (c). We also approximate the wave number spectrum of damped PICWs with a red circle at $\kprl\gtrsim \kprl^*$ in (a). }
    \label{fig:PICW_WKB}
\end{figure}

We now detail how the propagation and advection of PICWs in the inhomogeneous solar wind changes the properties of the waves and, inevitably, leads to their dissipation. As PICWs are driven unstable by the turbulent cascade, they will both propagate and be advected anti-Sunward. Let us assume the solar wind background is in steady-state. Then, the real part of the angular frequency of PICWs, measured in the heliocentric frame, will be constant with radius, {\em viz.}
\begin{equation}\label{eqn:PICW_WKB}
    \omega_{\rm r\odot}=\kprl U+\omegaP= \const.
\end{equation}
In the $U\gg \vA$ super-Alv{\'e}nic wind, ${\omegaP\le \kprl\vA\ll \kprl U}$, and so ${\omega_{\rm r\odot}=\omegaP+\kprl U\simeq \kprl U}$. For heliocentric distances much larger than the Alfv{\'e}n critical point, ${U\sim \const}$, so ${\kprl\sim\const}$. Thus, ${\kprl d_{\rm p}\sim n^{-1/2}\sim x}$, where $x=r/\Rsun$ is the heliocentric distance normalized to the solar radius $\Rsun$. In the sub-Alfv{\'e}nic wind, the same analysis, again taking the non-dispersive limit, yields ${\kprl\vA\sim\const}$, so ${\kprl d_{\rm p}\sim B^{-1}\sim x^2}$ at least for $x\gtrsim 3$. While the radial scaling of $U$ near the Alfv{\'e}n critical point somewhat complicates the preceding analysis, the general qualitative conclusion is clear: the normalized parallel wave number, $\kprl d_{\rm p}$, of PICWs increases as the waves travel away from the Sun.

Depicted diagrammatically, the radial evolution of $\omegaP/\Omega_{\rm p}$ and $\kprl d_{\rm p}$ is incredibly simple because the PICW cold plasma dispersion relation \eqref{eqn:wPICW} is invariant with heliocentric distance if $\omegaP$ and $\kprl$ are normalized by $\Omega_{\rm p}$ and $d_{\rm p}$, respectively. As a result, the normalized frequency and parallel wave number of a given PICW simply increase along the normalized dispersion relation as the waves are transported to larger radii. We depict this evolution with a black arrow in figure~\ref{fig:PICW_WKB}(c) and vertical black arrow in figure~\ref{fig:PICW_WKB}(a). As $\kprl d_{\rm p}$ increases, the resonant parallel velocity decreases, as depicted by the shallower red resonance line in figure~\ref{fig:PICW_WKB}(c). If PICW emission and absorption occur continuously across a range of heliocentric distances, then at any given heliocentric distance one will observe PICWs cooling protons with $\vprl<\vprl^*$ and heating protons with $\vprl^*<\vprl<0$, as shown in figure~\ref{fig:PICW_WKB}(b). This distinct heating signature was recently produced by a quasi-linear analysis of PICWs observed by \textit{PSP} during an encounter with a fast solar wind stream \citep{Bowen_2024}.

\section{Conclusion}\label{sec:conclusion}
In this work, we have shown that observations of heating by PICWs in the fast solar wind can be explained by a continuous process of PICW emission, propagation, and absorption that we call cyclotron breaking. The name `cyclotron breaking' comes as an analogy to water waves breaking on the shore. In the same way that water waves can be driven by a shoreward wind interacting with the surface of the water, PICWs can be driven by the turbulent cascade of anti-Sunward Alfv{\'e}n waves interacting with the proton distribution. Both waves then propagate in an inhomogeneous medium: the water waves in increasingly shallower water as they approach the shore, and PICWs in increasingly rarefied plasma as they are transported away from the Sun. As they propagate, the wavelength and frequency of both waves evolve such that the waves go from gaining in amplitude to dissipating their energy. In the case of water waves, the waveform itself undergoes nonlinear steepening as it moves up the shore, eventually reaching a point where the wave breaks and turbulently dissipates. PICWs, on the other hand, evolve from resonating with relatively fast protons, to which they are unstable, to resonating with relatively slow protons, on which they damp. While the PICW damping described by this process is purely linear, we nevertheless borrow the concept of wave breaking here as a metaphorical comparison.

Cyclotron breaking is reminiscent of the sweeping scenario, in which it is hypothesized that the Sun launches kHz-range ICWs \citep{Axford_1992} that damp as they propagate away from the Sun. The damping of these waves occurs at a heliocentric distance at which the wave frequency is comparable to the local cyclotron frequency, which is a radially decreasing function of heliocentric distance \citep{Schwartz_1981, Tu_1997}. The cyclotron frequency thus `sweeps' down through the wave power spectrum, damping out the waves first at high frequencies near the Sun and then lower frequencies as~$r$ increases. A difference between sweeping and cyclotron breaking is that in the sweeping scenario the PICWs are generated at the Sun, whereas in cyclotron breaking they are generated in the solar wind via quasilinear focusing of oblique ICWs that are produced locally by the turbulent cascade. Cyclotron breaking therefore mitigates many of the observational inconsistencies with the solar origin of ICWs in the sweeping model \citep{Cranmer_2000,Hollweg_2000,Cranmer_2001}.

The unique heating signature predicted by cyclotron breaking -- PICWs cooling relatively fast protons while heating relatively slow protons -- is consistent with heating signatures produced by quasi-linear analysis of PICW heating in the fast solar wind \citep{Bowen_2024}. The same analysis, however, also found that PICWs on average heat the solar wind. It is not yet established under what conditions cyclotron breaking can explain this result. Net heating will occur at a given heliocentric distance when the PICW power being absorbed is greater than the PICW power being emitted. If PICWs are driven unstable at a specific heliocentric distance, they will only begin to damp once they have propagated away from where they were driven unstable. It is intuitively clear that these waves can on average heat the solar wind at heliocentric distances larger than where they were emitted because there is no local PICW emission to offset the damping of the waves. 

For a process of continuous PICW emission and absorption more akin to what is observed in the solar wind \citep{Bowen_2024}, determining the balance between local heating and cooling at a given heliocentric distance is a non-trivial task. To do this properly, a more comprehensive modeling effort will need to be undertaken in which the radial evolution of $\vprl^*$, $\kresP^*$, and PICW wave amplitude are treated self-consistently. Broadly speaking, however, one can imagine that if the incident flux of PICWs with $\kprl>\kresP^*$ is sufficiently large -- and that their damping is sufficiently strong -- local heating will dominate local cooling. So long as the power emitted as PICWs decreases sufficiently quickly as a function of radius, this condition may well be satisfied. Indeed, \citet{Bowen_2024} found that the PICW heating rate decreased rapidly with heliocentric distance, ${\sim}x^{-a}$ with $4<a<5$, so it appears plausible that cyclotron breaking can explain observations of net heating by PICWs in the young, fast solar wind.

\section*{Acknowledgments}
We would like to thank Matt Kunz, Jono Squire, Michael Zhang, Kris Klein, Nikos Sioulas, Chris Chen,  Niranjana Shankarappa, Lynn Wilson, Toby Adkins, and the participants of the Parker Four conference, whose comments and conversations helped refine this work. This work was supported by NASA grants 80NSSC24K0171 and NNN06AA01C  and DOE award DE-SC0026201.

\appendix

\section{Proof that $\omegakr>\omegakprlr$ for all $(\kprl,\kprp)>0$}\label{apx:proof}
Let us prove that $\omegakr>\omegakprlr$ for all $(\kprl,\kprp)>0$. If $\omegakr>\omegakprlr$, then, noting that $\omegakr>0$ and $\omegakprlr>0$ by definition, $2\omegatilde_{k\rm r}^2>2\omegatilde_{\kprl\rm r}^2$, where $\omegatilde_{k\rm r}=\omegakr/\Omega_{\rm p}$ and $\omegatilde_{\kprl\rm r}=\omegakprlr/\Omega_{\rm p}$. Using the dispersion relation \eqref{eqn:wICW} and rearranging, 
the inequality becomes
\begin{equation}\label{eqn:proof1}
    b_{\rm O}-b_{\rm P}>\sqrt{d_{\rm O}}-\sqrt{d_{\rm P}},
\end{equation}
where $d_{\rm O/P}=b_{\rm O/P}^2-4c_{\rm O/P}$, $b_{\rm O}=b_{\rm P}+\kprptilde^2\left(1+\kprltilde^2\right)$, $b_{\rm P}=\kprltilde^2\left(2+\kprltilde^2\right)$, $c_{\rm O}=c_{\rm P}+\kprltilde^2\kprptilde^2$, $c_{\rm P}=\kprltilde^4$, $\kprltilde=\kprl d_{\rm p}$, and $\kprptilde=\kprp d_{\rm p}$. Using the above definitions, one can show that 
\begin{subequations}
\begin{align}
    d_{\rm O}&=d_{\rm P}+\kprptilde^4\left(1+\kprltilde^2\right)^2+2\kprptilde^2\kprltilde^4\left(3+\kprltilde^2\right)\label{eqn:proof_dO}\\
    d_{\rm P}&=\kprltilde^6\left(4+\kprltilde^2\right)\label{eqn:proof_dP}
\end{align}
\end{subequations}
Thus $b_{\rm O}>b_{\rm P}>0$ and $d_{\rm O}>d_{\rm P}>0$, so both sides of \eqref{eqn:proof1} can be squared with the inequality maintained. After some simplification, one finds
\begin{equation*}
    \sqrt{d_{\rm O}d_{\rm P}}>d_{\rm P} + \kprptilde^2\kprltilde^4\left(3+\kprltilde^2\right).
\end{equation*}
As before, both sides of the inequality are positive and can be squared with the inequality intact. After substituting \eqref{eqn:proof_dO}, a number of terms drop out and the inequality becomes
\begin{equation*}
    \kprp^4\left[d_{\rm P}\left(1+\kprltilde^2\right)^2-\kprltilde^8\left(3+\kprltilde^2\right)^2\right]>0.
\end{equation*}
Upon substituting \eqref{eqn:proof_dP} and simplifying, one is left with
\begin{equation}
    \kprptilde^4\kprltilde^6>0,
\end{equation}
which is indeed satisfied for all $(\kprl,\kprp)>(0,0)$. QED.

\bibliographystyle{jpp}

\bibliography{main}

\end{document}